\documentclass[journal=nalefd,manuscript=letter,layout=twocolumn]{achemso}
\pdfoutput=1
\usepackage{color}
\usepackage{dcolumn}
\usepackage{amssymb}
\usepackage{amsmath}
\usepackage{graphicx}
\usepackage[normalem]{ulem}
\usepackage[pdftex]{hyperref}
\hypersetup{colorlinks=true,citecolor=blue,linkcolor=red,urlcolor=blue}
\usepackage[all]{hypcap}
\usepackage{rotating}

\author{Roberto Guerra} \email{roberto.guerra@unimi.it}
  \affiliation{Center for Complexity and Biosystems, Department of Physics, University of Milan, 20133 Milan, Italy}
  \affiliation{International School for Advanced Studies (SISSA), Via Bonomea 265, 34136 Trieste, Italy}

\author{Itai Leven}
  \affiliation{Department of Physical Chemistry, School of Chemistry,
  The Raymond and Beverly Sackler Faculty of Exact Sciences and The
  Sackler Center for Computational Molecular and Materials Science,
  Tel Aviv University, Tel Aviv 6997801, Israel}

\author{Andrea Vanossi}
  \affiliation{CNR-IOM Democritos National Simulation Center, Via Bonomea 265, 34136 Trieste, Italy}
  \affiliation{International School for Advanced Studies (SISSA), Via Bonomea 265, 34136 Trieste, Italy}

\author{Oded Hod}
  \affiliation{Department of Physical Chemistry, School of Chemistry,
  The Raymond and Beverly Sackler Faculty of Exact Sciences and The
  Sackler Center for Computational Molecular and Materials Science,
  Tel Aviv University, Tel Aviv 6997801, Israel}

\author{Erio Tosatti} 
  \affiliation{International School for Advanced Studies (SISSA), Via Bonomea 265, 34136 Trieste, Italy}
  \affiliation{CNR-IOM Democritos National Simulation Center, Via Bonomea 265, 34136 Trieste, Italy}
  \affiliation{The Abdus Salam International Centre for Theoretical Physics (ICTP), Strada Costiera 11, 34151 Trieste, Italy}

\title{The Smallest Archimedean Screw:\\ Facet Dynamics and Friction in Multi-Walled Nanotubes}


\begin{document}

\begin{abstract}
We identify a new material phenomenon, where minute mechanical
manipulations induce pronounced global structural reconfigurations in
faceted multi-walled nanotubes. This behavior has strong implications
on the tribological properties of these systems and may be the key to
understand the enhanced inter-wall friction recently measured for
boron-nitride nanotubes with respect to their carbon counterparts.
Notably, the fast rotation of helical facets in these systems
upon coaxial sliding may serve as a nanoscale Archimedean screw for
directional transport of physisorbed molecules.
\end{abstract}

\section{Introduction}
Nanotubes\cite{Iijima1991,Tenne1992,Rubio1994,Chopra1995,Cohen2010}
form a paradigmatic family of quasi-one-dimensional materials playing
a central role in the design of many nano-electro-mechanical
systems.\cite{Zheng2002,Joselevich2006,Stampfer2006,Jensen2006,Nagapriya2008,Lucas2009,Garel2012,Chiu2012,Levi2013,Volder2013,Arash2014,Garel2014,Levi2015,Divon2017}
Traditionally, they are perceived as miniature cylinders of nanoscale
circular cross-section. Nevertheless, if the chirality of neighboring
shells within a concentric multi-walled nanotube is correlated,
extended circumferential facets may
form.\cite{Liu1994,Gogotsi2000,Zhang2003,Zhang2005,Celik-Atkas2005-1,Golberg2007,Leven2016}
The resulting polygonal cross section induces geometric inter-wall
locking that can considerably enhance their mechanical
rigidity.\cite{Garel2012}

Despite their remarkable structural similarity, faceting is
more commonly observed in multi-walled boron-nitride nanotubes
(MWBNNTs)\cite{Celik-Atkas2005-1,Golberg2007,Garel2012,Garel2014}
than in their carbon counterparts
(MWCNTs).\cite{Liu1994,Gogotsi2000,Zhang2003,Zhang2005} This can be
attributed to three important factors: (i) Stronger long-range
dispersive attractive interactions exhibited by the
former\cite{Zheng2012,Hsing2014,Leven2014}
that provide higher inter-wall adhesion thus favoring
facet formation; (ii) Softer ZA modes of {\it h}-BN\cite{wirtz2003}
that allow for sharper vertices thus promoting the formation of wider
planar facet regions; and (iii) Higher inter-wall chiral angle
correlation exhibited by MWBNNTs over
MWCNTs\cite{Golberg1999,Golberg2000,Zuo2003,Li2003,Celik-Atkas2005-1,
Celik-Atkas2005-2,Hashimoto2005,Koziol2005,Xu2006,Ducati2006,Hirahara2006,
Gao2006,Guan2008,Schouteden2013}
that induces extended lattice registry patterns between adjacent tube
shells and dictates the nature of the
facets.\cite{Kolmogorov2005,Leven2016} The latter is mainly due to the
additional inter-wall electrostatic interactions between the partially
charged ionic centers in the hetero-nuclear BNNT network. While being
relatively weak locally,\cite{Lennard-Jones1928,Hod2012} when summed
over extended commensurate facet regions these Coulomb interactions
can foster energetic stabilization.

Similar to macroscopic objects, the strain developing within the
hexagonal lattice of nanotubes under small external mechanical
manipulations is proportional to the applied stress. Due to their
exceptionally high rigidity this usually leads to minor structural
deformations. In the present study, however, we discover a new
material phenomenon, occurring in faceted double-walled nanotubes
(DWNTs), where minute mechanical manipulations induce pronounced
global superstructure reconfiguration. For monochiral DWNTs that
exhibit axially aligned facets\cite{Leven2016} even the slightest
inter-wall rotation induces significant circumferential facet
revolution, and minor inter-wall telescoping can lead to complete
unfaceting. Similar manipulations applied to bichiral DWNTs result in
global screw-like motion of their elongated helical facets reminiscent
of an Archimedean screw. Importantly, these superstructure
evolutions under coaxial sliding open new collective energy dissipation channels that
enhance inter-wall dynamic friction. This, in turn, suggests that the
relative abundance of faceting in MWBNNTs plays a central role not
only in their enhanced torsional stiffness\cite{Garel2012} but also in
the significantly higher inter-wall friction that they exhibit with
respect to MWCNTs.\cite{Nigues2014}

\section{Facet Superstructure Reconfiguration}
To demonstrate the phenomenon of superstructure reconfiguration we
consider a set of four representative double-walled BNNTs (DWBNNTs)
including the achiral armchair (104,104)@(109,109) and zigzag
(180,0)@(188,0) DWBNNTs that present axial facets; the
bichiral (70,70)@(77,74) system, whose small inter-wall chiral angle difference of
$\Delta\Theta$\,=\,0.657$^\circ$ induces helical facets;
and the achiral mixed (179,0)@(108,108) DWBNNT ($\Delta\Theta$\,=\,30$^\circ$)
that does not form circumferential facets.\cite{Leven2016}
Here, the notation $(n_1,m_1)$@$(n_2,m_2)$ represents a $(n_1,m_1)$ inner
tube concentrically aligned within an outer $(n_2,m_2)$ tube, where $n_i$
and $m_i$ are the corresponding tube indices.\cite{Saito-Dresselhauses}

\begin{figure*}[tb]
  \centering
  \includegraphics[draft=false,width=0.6\textwidth]{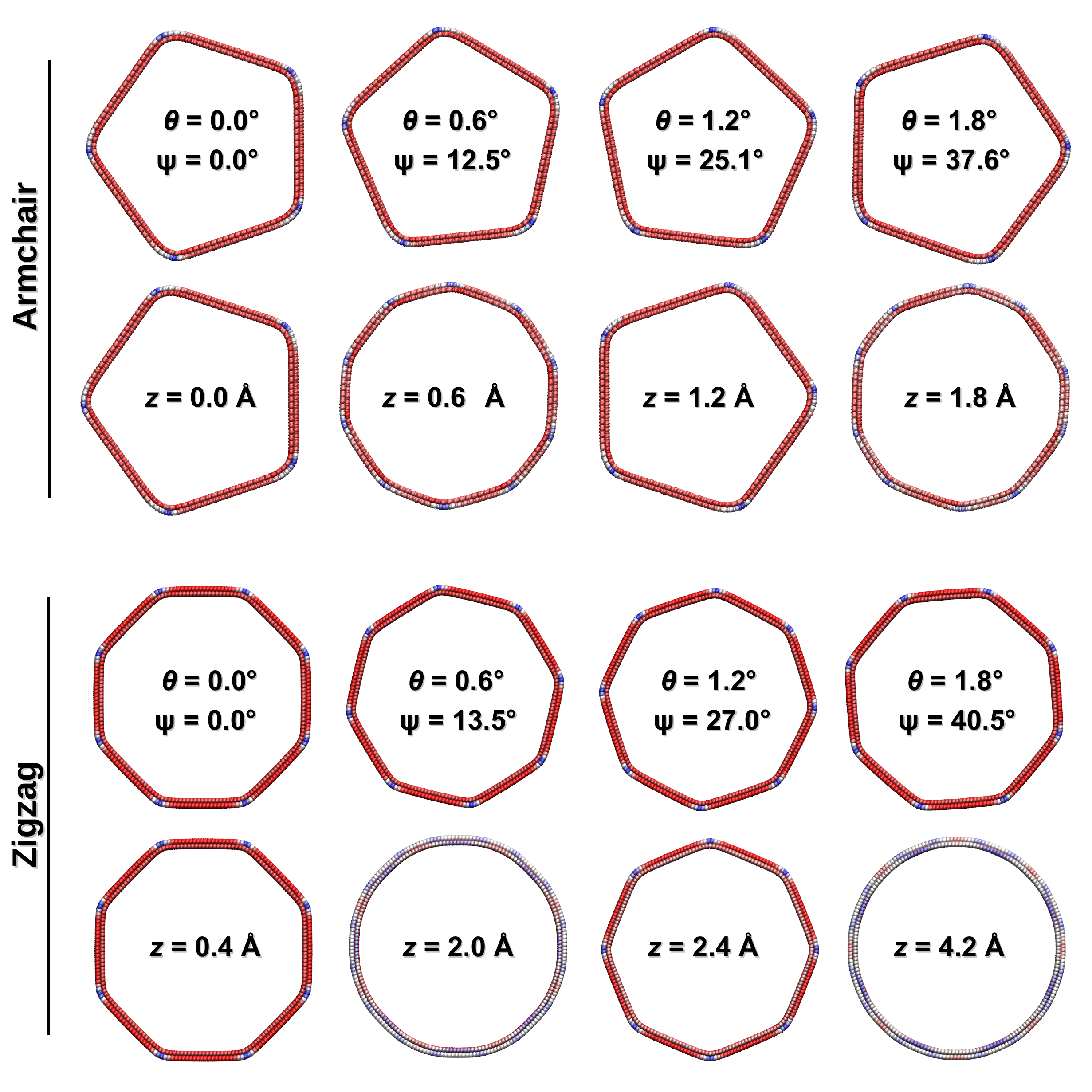}
  \caption{
    \small {\bf Achiral faceting} -- Cross sectional view of the
    achiral armchair (top rows) and zigzag (bottom rows) DWBNNTs
    during a full coaxial sliding cycle. Each row shows configurations
    at increasing relative angular ($\theta$) and axial ($z$)
    difference between the outer and inner walls. Continued motion beyond the
    domain considered herein results in periodic repetitions of the presented
    structures. $\Psi$ indicates the corresponding facet rotation angle.
    Intermediate configurations during the adiabatic pull-out process appear
    in Supporting Information Movie 1. Red, white, and blue colors indicate low,
    average, and high atomic interlayer energy, respectively.
    }
    \label{fig.facets_achiral}
\end{figure*}

Focusing first on inter-wall rotations of the achiral systems, we
perform a set of constrained energy minimizations starting from a
circular DWBNNT configuration and relaxing the geometry at several
fixed inter-wall angular orientations ranging from 0$^{\circ}$ to
2$^{\circ}$. Fig.~\ref{fig.facets_achiral} presents the corresponding
relaxed structures of the armchair (104,104)@(109,109) (first row) and
the zigzag (180,0)@(188,0) (third row) DWBNNTs. As is
evident
from the figure, the angular orientation of the facets shows strong
dependence on the inter-wall rotation angle. In the armchair case,
which presents an optimal structure of pentagonal cross section,
a dramatic 41.8$^{\circ}$
revolution of the facet superstructure is obtained for every nominal
inter-wall rotation of 2$^{\circ}$. Similarly, the octagonal
circumferential superstructure of the zigzag case
revolves by as much as 45$^{\circ}$ at a similar inter-wall rotation of
2$^{\circ}$.

All the more pronounced structural variations arise in response to
inter-wall telescoping. For the armchair DWBNNTs considered (second row
of Fig.~\ref{fig.facets_achiral}) the number of facets doubles from 5
to 10 and their angular orientation rotates by 18$^\circ$ upon
inter-wall telescoping of 1.25\,\AA.
Notably, for the zigzag DWBNNT (lowest row of Fig.~\ref{fig.facets_achiral})
almost complete unfaceting is observed upon an axial shift of merely $\simeq$\,1.7\,\AA.
The entire structural variation progression obtained during an adiabatic pull-out
of 4.2\,\AA\ is reported in Supporting Information Movie 1.

\begin{figure*}[tb]
  \centering
  \includegraphics[draft=false,width=0.7\textwidth]{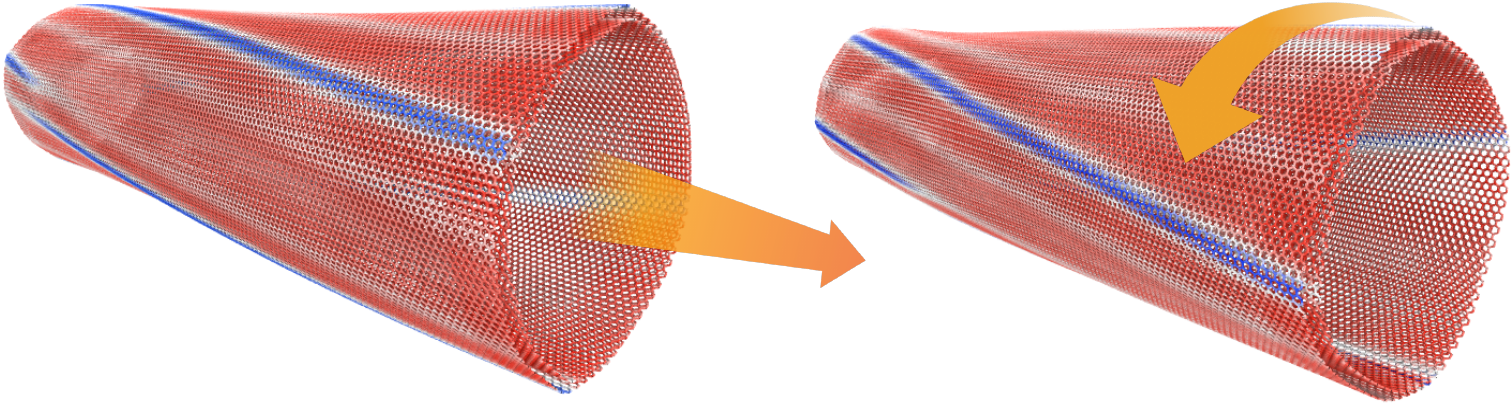}
  \caption{
    \small {\bf Bi-chiral DWNT facet rotation} -- Perspective view of
    the bi-chiral (70,70)@(77,74) DWBNNT at two inter-wall configurations
    $\theta$/$z$ of 0.2$^\circ$/2.4\,\AA\ (left) and 0.2$^\circ$/3.2\,\AA\ (right).
    These correspond to configurations close to maximum and minimum potential energy,
    respectively (see top-right panel of Fig.~\ref{fig.PES}).
    Blue and red atom false coloring represents high and low interlayer energy,
    respectively. Facet dynamics during a pull-out simulation at an inter-wall
    velocity of 0.01\,\AA/ps is reported in Supporting Information Movie 3.
  }
  \label{fig.facets_chiral}
\end{figure*}

The most remarkable structural response is exhibited by the bi-chiral
(70,70)@(77,74) DWBNNT. The chiral facets appearing in this system
couple the translational and rotational degrees of freedom. Hence,
inter-wall telescoping induces global rotation of the entire helical
superstructure reminiscent of an Archimedean screw (see
Fig.~\ref{fig.facets_chiral} and Supporting Information Movie 3).
This represents the smallest device exhibiting
unidirectional helical motion that may be utilized as a nanoscale
arterial thoroughfare for molecular transport.

On the contrary, the achiral mixed zigzag@armchair (179,0)@(108,108)
DWBNNT that possesses the maximal inter-wall chiral angle difference of
$\Delta\Theta$\,=\,30$^\circ$ presents a featureless circular cross section
(not shown) regardless of the inter-wall position.

\section{Potential Energy Landscapes}
The significant superstructure variations described above are expected
to have distinct manifestation in the mechanical and tribological
characteristics of faceted nanotube structures. To evaluate these, we
compare, in Fig.~\ref{fig.PES}, the potential energy surface (PES)
for inter-wall rotation and telescoping of the faceted DWBNNTs
(for the carbon counterpart see Supporting Information Fig.~S1)
considered with those of their circular cross-section counterparts
(see Structures and Methods section for technical details regarding the calculation).

Focusing first on the armchair (104,104)@(109,109) DWBNNT (left column
in Fig.~\ref{fig.PES}) we find, as expected, that the potential
energy corrugation for inter-wall rotations of the circular
configuration is very small (2.7$\cdot$10$^{-4}$\,meV/atom).
This results from the inter-wall curvature difference that
induces circumferential incommensurability between the hexagonal lattices
of the two nanotube shells.\cite{KCrespi2000} Interestingly, the faceted
configuration maintains a smooth inter-wall rotation energy landscape
(corrugation of 3.0$\cdot$10$^{-5}$\,meV/atom) indicating
in addition that pure adiabatic facet
reorientation is practically a barrierless process. On the contrary,
even for the circular configuration, inter-wall telescoping is
associated with potential energy variations that follow the mutual
hexagonal lattice periodicity, $p_z$, along the zigzag axial direction
of the two nanotube walls ($p_z$\,=\,$l\sqrt{3}$\,$\simeq$\,2.498\,\AA,
where $l$\,=\,1.442\AA\ is the equilibrium BN bond length).
Notably, for the faceted configuration the amplitude of these variations
is an order of magnitude larger (0.84\,meV/atom) than that of the circular
system (0.074\,meV/atom), as is also reflected by the energy landscapes of Fig.~\ref{fig.PES}.
This is due to the unfaceting and refaceting
restructuring sequence occurring during the pull-out process
(see Supporting Information Movie 2).
At the faceted configuration the average inter-wall
distance reduces from its circular cross-section value of 3.44\,\AA\
-- a value geometrically determined by the lattice indices of the two walls
and by $l$ -- to an optimal inter-facet separation of 3.27\,\AA,
matching the equilibrium {\it h}-BN bilayer interlayer distance
of our interatomic potential (see Structures and Methods section).
Hence, the overall inter-wall steric repulsion
increases with respect to the unfaceted configuration resulting in
higher telescoping PES barriers.

The zigzag (180,0)@(188,0) DWBNNT exhibits similar behavior with
smooth inter-wall rotation (1.4$\cdot$10$^{-4}$ and 6.1$\cdot$10$^{-5}$\,meV/atom for the
circular and faceted configurations, respectively) and a corrugated
telescoping PES (middle column in Fig.~\ref{fig.PES}). The latter now
follows the periodicity, $p_a$, of the armchair axial direction of the two
nanotube walls ($p_a$\,=\,3\,$l$\,=\,4.326\AA).
Unlike the armchair DWBNNT case discussed above, here the circular zigzag DWBNNT
configuration presents considerably higher corrugation (8.4\,meV/atom)
than the faceted one (1.4\,meV/atom). This results
from the fact that the inter-wall distance in the frustrated circular
system, 3.18\,\AA, is smaller than the optimal value. Upon facet
formation the inter-facet distance now increases to a nearly optimal
value of 3.25\,\AA.
This, in turn, results in lower barriers along the unfaceting and
refaceting sequence obtained throughout the pull-out process.

An overall lower PES corrugation is presented by the circular bichiral
(70,70)@(77,74) DWBNNT (right columns of Fig.~\ref{fig.PES}) with
relatively smooth telescoping and inter-wall rotation energy profiles
(2.3$\cdot$10$^{-8}$ and 1.3$\cdot$10$^{-4}$\,meV/atom, respectively).
This mainly results from the fact that the inter-wall
distance at this configuration, 3.78\,\AA, is considerably larger
than the equilibrium value. Similar to the case of the armchair system
described above, the appearance of facets effectively reduces the
inter-facet distance toward the equilibrium value resulting in an
increase of the PES corrugation. Nevertheless, while the translational
and rotational degrees of freedom remain decoupled in the achiral
systems that present axial facets, here they are strongly coupled by
the helical facets as demonstrated by the tilted (rather than vertical
or horizontal) PES ridges.

\newcommand*{\xdash}[1][3em]{\rule[0.5ex]{#1}{0.55pt}}
\begin{figure*}[tb]
  \centering
  \hspace{7mm }\xdash[3.em]~~{\sf Armchair}~~\xdash[3.em]
  \hspace{6mm}\xdash[3.em]~~{\sf Zigzag  }~~\xdash[3.em]
  \hspace{6mm}\xdash[3.em]~~{\sf Bichiral}~~\xdash[3.em]\\
  \begin{turn}{90}
    ~ \xdash[1.3em]~~{\sf Relaxed}~~\xdash[1.3em]
  \end{turn}
  \hspace*{2mm}
  \includegraphics[draft=false,width=0.31\textwidth]{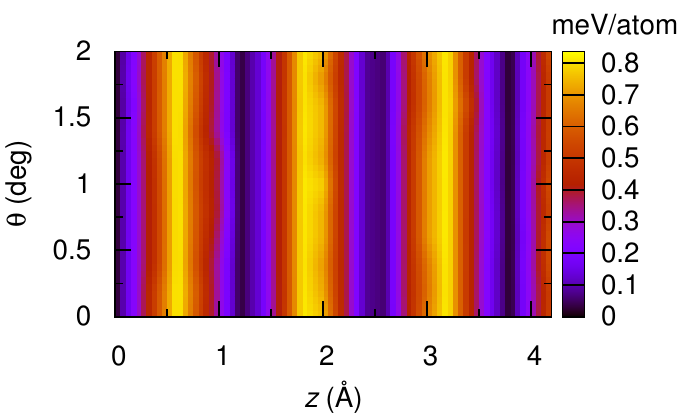}
  \includegraphics[draft=false,width=0.31\textwidth]{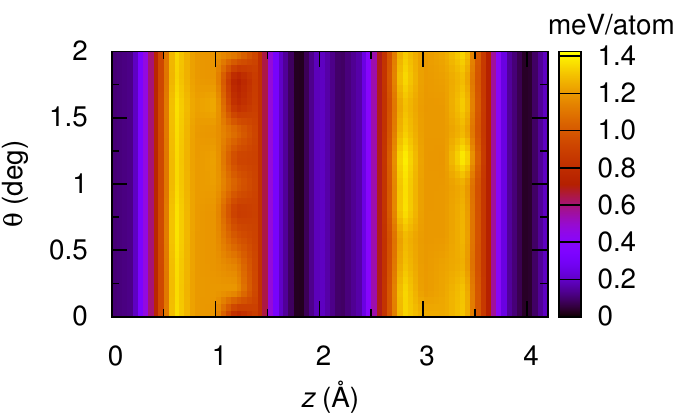}
  \includegraphics[draft=false,width=0.31\textwidth]{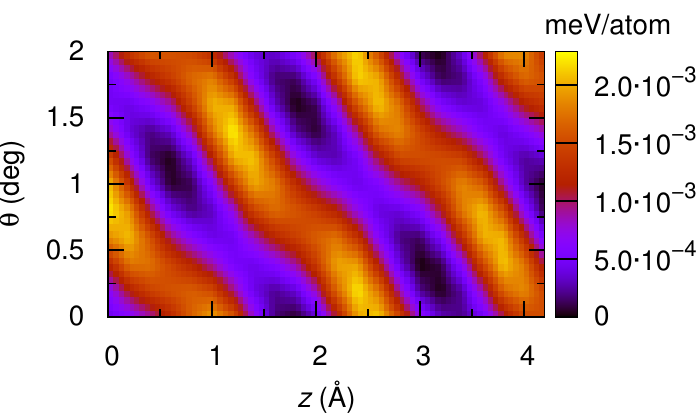}\\
  \begin{turn}{90}
    ~~ \xdash[1.em]~~{\sf Cylindric}~~\xdash[1.em]
  \end{turn}
  \hspace*{2mm}
  \includegraphics[draft=false,width=0.31\textwidth]{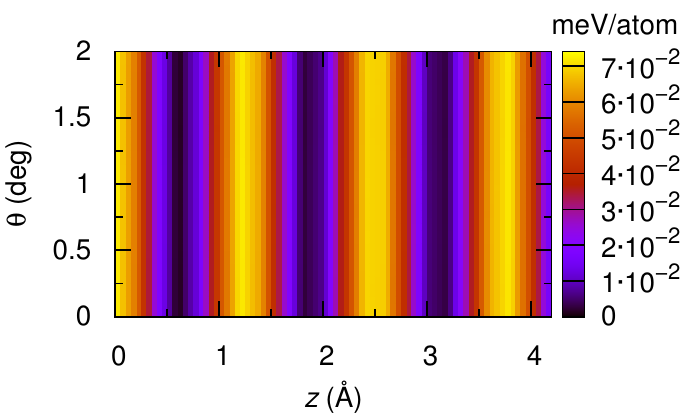}
  \includegraphics[draft=false,width=0.31\textwidth]{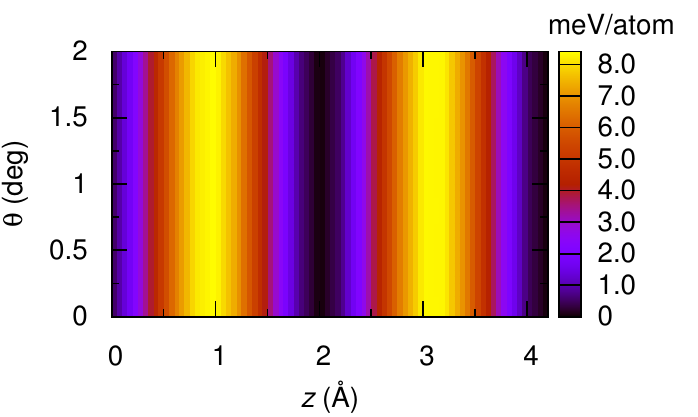}
  \includegraphics[draft=false,width=0.31\textwidth]{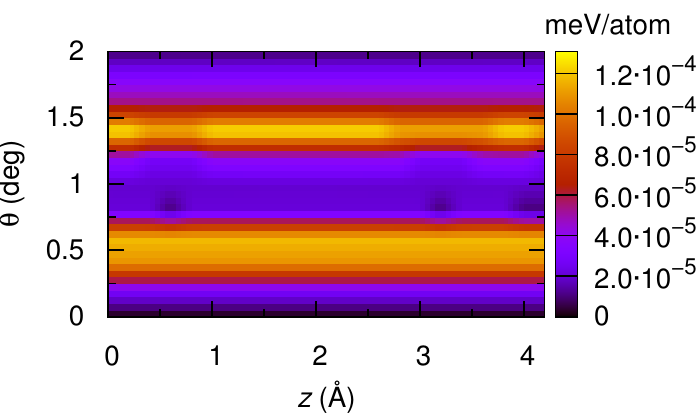}\\
\caption{\small {\bf PES maps} -- Potential energy surface maps of the considered
  armchair (left column), zigzag (center column), and bi-chiral
 (right column) DWBNNTs for relaxed (top panels) and cylindric (bottom
 panels) configuration.}\label{fig.PES}
\end{figure*}

Interestingly, the achiral mixed (179,0)@(108,108) DWBNNT has an
inter-wall distance of 3.21\,\AA, comparable to that of the zigzag
(180,0)@(188,0) DWBNNT and smaller than the equilibrium value. One
might therefore conclude that the two systems should present similar
PES corrugation. Nevertheless, the former presents a completely flat
translational-rotational PES for both the constrained circular and the
fully relaxed configurations (not shown). This may be attributed to
the incommensurability of the two hexagonal lattices in both the axial
and circumferential directions obtained at the maximal inter-wall
chiral angle misfit of 30$^\circ$.

\section{Inter-wall Static Friction}
A twofold effect of circumferential faceting on the inter-wall PES of
DWBNNTs is thus found: (i) Facet restructuring during inter-wall
displacements results in inter-wall distance variations that may
increase or decrease PES corrugation depending on the corresponding
distance within the unfaceted system; (ii) Helical facets, appearing
in bichiral DWBNNTs, couple the translational and rotational degrees
of freedom. The immediate physical manifestation of these effects is
expected to appear in the static inter-wall friction exhibited by the
DWNT.

The static friction force is defined as the minimal force required to
initiate relative motion between the two nanotube walls that are
initially interlocked in a (local) free-energy minimum.
Despite the general non-uniformity of real telescopic sliding,
also depending on the pulling mode, static friction may, in the low
temperature limit ($T$\,$\rightarrow$\,0\,K), be estimated from the
inter-wall telescoping-rotation PES by evaluating the energy barrier
required to lift the interface out of the equilibrium state. To this
end, we plot the energy variations during adiabatic axial inter-wall
pull-out and rotation and fit them to a sinusoidal curve of the form
$E(z)$\,=\,$(E_c/2)\sin(2\pi z/\Delta z)$ (See Supporting Information Fig.~S2).
The static friction is then extracted from the maximal
derivative of the fitted curve given by
$F_s$\,=\,${\pi}E_c/{\Delta}z$.
A summary of the obtained PES corrugation and the corresponding static
friction values appears in Supporting Information Tab.~S1.

As may be expected, for the achiral armchair and zigzag systems the
static friction force required to initiate inter-wall rotational
motion is negligible compared to that necessary to trigger telescopic
sliding for both the circular and the faceted configurations. The
pull-out static friction force of the armchair DWBNNT at the circular
geometry is 0.17\,meV/\AA\ per atom, much lower than
the corresponding value of the zigzag system (11.8\,meV/\AA).
As discussed above, in the relaxed configuration the inter-facet distance
approaches the equilibrium value in both systems resulting in similar
friction forces of 1.91 and 2.04 meV/\AA\ per atom for the armchair
and zigzag DWBNNTs, respectively.

The bichiral system in its circular geometry presents a negligible
static friction force for axial shifts (3.8$\cdot$10$^{-8}$\,meV/\AA\ per
atom), while a larger value, yet considerably smaller than the
characteristic forces exhibited by the achiral systems, is obtained
for inter-wall rotations (2.2$\cdot$10$^{-4}$ meV/\AA\ per atom). At the
faceted configuration the static friction forces for both telescoping
and rotation increase yielding values of about 3.5$\cdot$10$^{-3}$
and 2.5$\cdot$10$^{-3}$ meV/\AA\ per atom, respectively.
It is clear from the upper right panel of Fig.~\ref{fig.PES} that a
combined rotation and telescoping displacement path which follows the
facet helicity will result in a considerably lower static friction force.
For the mixed achiral DWBNNT considered that, as mentioned above,
exhibits completely flat rotation-telescoping PES maps for both the
circular and relaxed (unfaceted) geometries, we could not extract any
meaningful static friction force values.

\vspace{2mm}
\begin{figure*}[tb]
  \centering
  \hspace{7mm}\xdash[3em]~~{\sf Armchair}~~\xdash[3em]
  \hspace{6mm}\xdash[3em]~~{\sf Zigzag  }~~\xdash[3em]
  \hspace{6mm}\xdash[3em]~~{\sf Bichiral}~~\xdash[3em]\\
  \vspace*{2mm}
  \begin{turn}{90}

    ~~ \xdash[1.em]~~{\sf DWBNNTs}~~\xdash[1.em]
  \end{turn}
  \hspace*{2mm}
  \includegraphics[draft=false,width=0.31\textwidth]{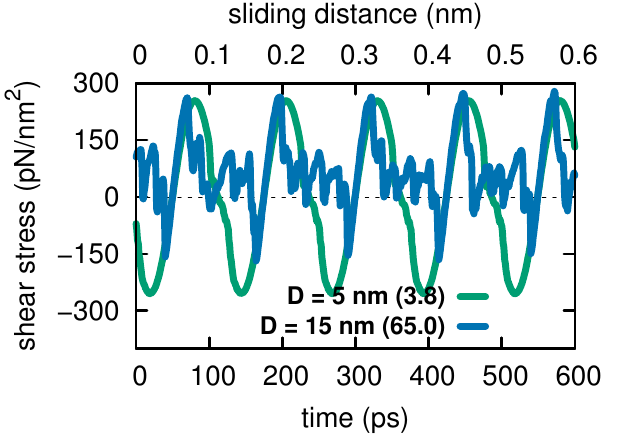}
  \includegraphics[draft=false,width=0.31\textwidth]{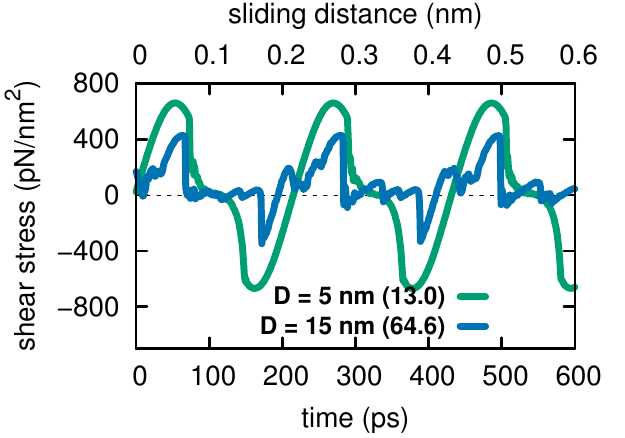}
  \includegraphics[draft=false,width=0.31\textwidth]{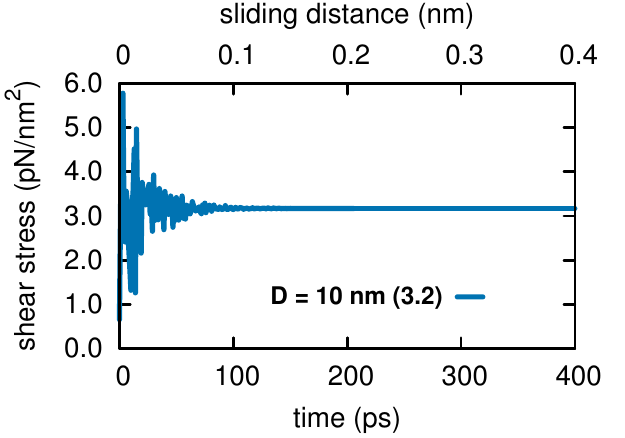} \\
  \begin{turn}{90}
    ~~ \xdash[1.em]~~{\sf DWCNTs}~~\xdash[1.em]
  \end{turn}
  \hspace*{2mm}
  \includegraphics[draft=false,width=0.31\textwidth]{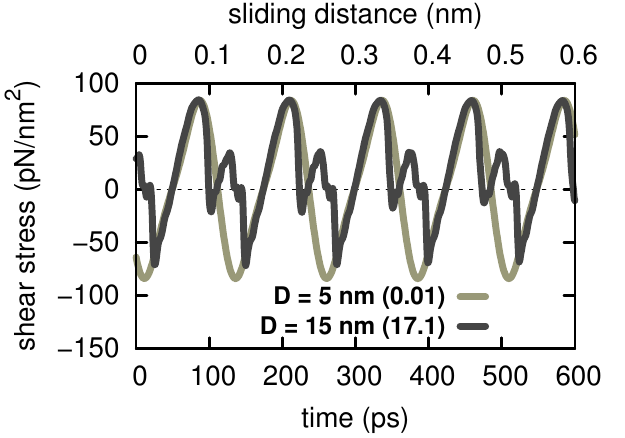}
  \includegraphics[draft=false,width=0.31\textwidth]{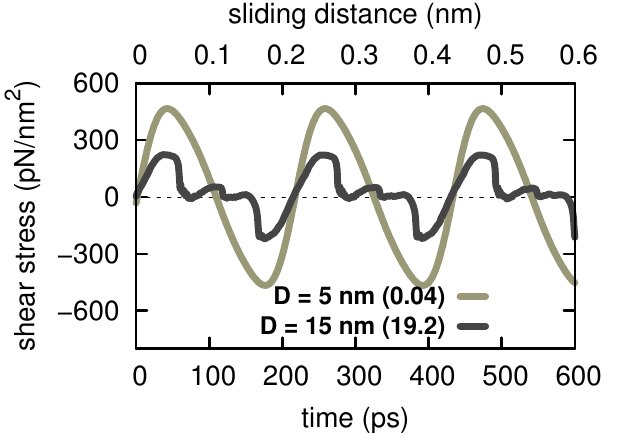}
  \includegraphics[draft=false,width=0.31\textwidth]{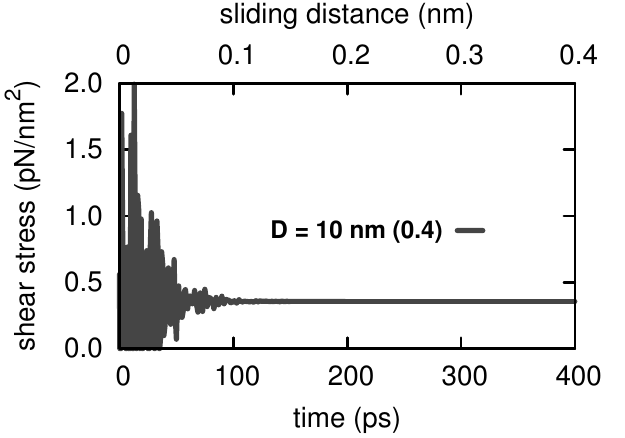} \\
  \caption{\small {\bf Dynamic friction} -- Instantaneous friction force per unit
    area (shear stress), calculated for inter-wall telescopic motion of
    armchair (left panel), zigzag (center panel), and bi-chiral (right panel) DWBNNTs (upper panels)
    and DWCNTs (lower panels) of diameter $D$ at a pull-out velocity of 0.01\,\AA/ps.
    For the bi-chiral case the initial transient dynamics is also shown.
    The average steady-state friction force values are reported in brackets in units of pN/nm$^2$.}\label{fig.friction}
\end{figure*}

\section{Dynamic Friction}
Not only do the superstructure reconfigurations described above impact
the static nanotube inter-wall friction but they provide a key
to understanding the surprisingly high inter-wall dynamic friction
recently measured for MWBNNTs with respect to their carbon
counterparts.\cite{Nigues2014} The underlying mechanism relates to the
fact that the facet superstructural collective degrees of freedom introduce
auxiliary energy dissipation routes that enhance dynamic friction.
This is true for both translational and rotational inter-wall motion
even when the latter presents negligible PES corrugation and static
friction forces.

To quantify these effects we performed fully atomistic molecular
dynamics inter-wall sliding simulations (see Structures and Methods
section for details) of the DWBNNTs considered. When following the
structural variations occurring during the telescopic pull-out of the
armchair and zigzag DWBNNTs' inner shells at a relative velocity of
0.01\,\AA/ps we observe a full unfaceting and refaceting superstructure
cycle, superposed on asymmetric deformations induced by inertial effects
(see Supporting Information Movie 2). For the bi-chiral DWBNNT we find that
telescopic motion, at the same relative axial velocity, induces
circumferential rotation of the helical facets with an angular
velocity of about 0.24\,$^{\circ}$/ps
(evaluated from the simulated time evolution in Supporting Information Movie 3).
This corresponds to a linear superstructure surface velocity of
$\approx$\,0.2\,\AA/ps (assuming an average tube diameter of
$\approx$\,10\,nm), which is more than 20 times faster than the
applied axial velocity.

Following Newton's first law, we define the instantaneous dynamic
friction force as the force required to maintain a constant velocity
relative inter-wall sliding motion. To allow for comparison between
DWNTs of different diameters we extract the shear stress by
normalizing the calculated forces to the nominal
surface contact area. In Fig.~\ref{fig.friction} the temporal
shear stress traces obtained during constant velocity inner
shell pull-out (see Structures and Methods section) are reported.
We start by considering the achiral armchair (104,104)@(109,109)
and zigzag (180,0)@(188,0) DWBNNTs (blue lines in the upper left
and middle panels, respectively).
To evaluate the effect of facet reconfiguration on the dynamic friction
force we perform reference calculations on narrow armchair (31,31)@(36,36)
and zigzag (55,0)@(63,0) DWBNNTs green lines in the upper left and middle
panels, respectively that are below the critical diameter for facet
formation.\cite{Leven2016,Garel2012}

Due to their axial inter-wall translational symmetry, the achiral
DWNTs present periodic dynamic friction force variations with large
peak values reflecting increased interfacial
commensurability. Interestingly, the overall amplitude variations of
the shear stress traces of the faceted DWBNNTs are comparable to those
of the narrower circular systems. Nevertheless, while the circular
systems present a nearly-sinusoidal smooth behavior, the faceted
DWBNNTs show a complex pattern of rapid force fluctuations with clear
asymmetry between the positive and negative shear stress regions. This
is a clear manifestation of the effects of superstructure
reconfigurations occuring during the pull-out dynamics in the presence
of facets. As a consequence, the dynamic friction force, evaluated as
the time averaged shear stress over an integer number of periods, is
found to be 5-17 times larger in the faceted achiral DWBNNTs than in
the circular systems studied.

We may therefore conclude that faceting which, as discussed above, is
considerably more prevalent in MWBNNTs than in MWCNTs, may be
responsible for the enhanced friction measured for the former.
To understand how the inter-wall friction of the less abundant faceted
MWCNTs compares to that of their BNNT counterparts, we have repeated
our calculations for the corresponding achiral DWCNTs (see lower panels of
Fig.~\ref{fig.friction}). Similar to the case of DWBNNTs, the
circular achiral DWCNTs show a much smoother and more symmetric
shear stress trace (see light-grey lines in the lower left and lower
middle panels of Fig.~\ref{fig.friction}) resulting in considerably
smaller dynamic friction forces than the faceted achiral systems
(dark-grey lines). Interestingly, even for the latter, the kinetic
friction force extracted is smaller by a factor of 3.4-3.8 than that
of the corresponding faceted DWBNNTs with the force-field parameters
used herein (see Structures and Methods section).
Importantly, this is true also for the zigzag (180,0)@(188,0) DWNTs considered,
where the PES corrugation of the BN based system was found to be comparable to
that of its carbon counterpart (Figs.~\ref{fig.PES} and S1).

For the bi-chiral (70,70)@(77,74) DWBNNT considered no periodic
kinetic friction force variations are observed (see upper right panel
of Fig.~\ref{fig.friction}). Furthermore, following some initial
transient dynamics, smooth steady-state sliding motion with
nearly-constant drag is obtained. This can be attributed to the
reduced inter-wall commensurability and PES corrugation in this system
(see right panels of Fig.~\ref{fig.PES}). Consequently, the average
dynamic friction force recorder in this case ($\sim$3.2\,pN/nm$^2$) is
20-fold times smaller than that of the faceted achiral
systems. Nevertheless, it remains nearly an order of magnitude larger
than the value measured for the corresponding bi-chiral DWCNT
(0.4\,pN/nm$^2$, see lower right panel of Fig.~\ref{fig.friction})
and a factor of 80-320 larger than the kinetic friction measured for
the achiral circular DWCNTs considered.

Finally, we study the velocity dependence of the interlayer sliding friction of
DWBNNTs and DWCNTs in the range of 0.2-1.0\,m/s (see Supporting Information Fig.\ S5).
Our results show nearly linear increase of the friction force with
the sliding velocity at the velocity range considered.
For the axially commensurate armchair DWNTs the friction extrapolates to a finite
value at zero velocity. This can be attributed to the finite static friction
exhibited by these systems. For the incommensurate bi-chiral DWNTs the friction
extrapolates to zero at vanishing interwall sliding velocity. This is in line
with the experimental observation of viscous inter-wall telescopic motion in
multi-walled NTs, where sliding is expected to occur at the weakest
incommensurate interface. The calculated inter-wall friction forces in
both DWCNTs considered are found to be weakly dependent on the sliding velocity and
are consistently lower than those obtained for the corresponding DWBNNTs. This
further supports the experimental observations of increased inter-wall friction
in MWBNNTs over MWCNTs.\cite{Nigues2014}

\section{Conclusions}

The resulting screw-like motion of the faceted helical pattern establishes the smallest
realization of an Archimedean screw with the potential to achieve directional transport
of weakly adsorbed molecules along the surface of the tube.

We note that the super-structure variations discussed above may be viewed as the nanotube
analogues of the soliton-like motion of moir\'e patterns occurring in sliding incommensurate
planar interfaces.\cite{Amidror2009} Nevertheless, due to geometric frustration in the tubular
configuration, the extended circumferential registry patterns result in considerably larger
structural deformations. The latter exhibit much richer dynamic behavior with marked influence
on the mechanical, tribological, and electronic properties of the system.

The motion of such collective degrees of freedom opens new dissipative channels that enhance
dynamic friction beyond the excitation of localized phonon modes. Since faceting is more commonly
observed in MWBNNTs than in their carbon counterparts this rationalizes recent experimental findings
showing that the former exhibit an order of magnitude larger dynamic friction.\cite{Nigues2014}
Furthermore, even when compared to the less abundant case of faceted DWCNTs, the BN systems exhibit
3--8 times larger dynamic friction forces. Hence, when designing smooth nanoscale bearings one should
resort to unfaceted MWCNTS\cite{KCrespi2000} whereas if torsional and axial rigidities are desired
facetted MWBNNTs should be the material of choice.\cite{Garel2012,Nigues2014}

Finally, several other, more speculative but highly intriguing, consequences of the striking facet
evolutions discussed herein can be envisioned. First, we have shown that facet dynamics strongly
depends on the relative chirality of adjacent nanotube walls. Therefore, the inter-wall pulling
force trace should encode information about the identity of the various tube shells. This, in turn,
opens new opportunities for novel material characterization techniques that may provide access to
the specific sequence of chiralities of successive nanotube walls. Furthermore, electronic effects,
not discussed herein, may also exhibit unexpected behavior. Specifically, surface states that typically
localize at sharp edges, such as the circumferential vertices of the polygonal cross-section, may also
be pumped along the surface of nanotubes in an Archimedean manner.

\cleardoublepage
\section*{Structures and Methods}

DWNTs can have inner and outer walls that are either zigzag (ZZ),
amrchair (AC), or Chiral (Ch). In the present study four types of
carbon and boron nitride (BN) DWNTs have been considered including the
achiral AC@AC (104,104)@(109,109) and ZZ@ZZ (180,0)@(188,0) systems;
the mixed achiral ZZ@AC (179,0)@(108,108); and the bi-chiral AC@Ch
(70,70)@(77x74). Here, the notation $(n_1,m_1)$@$(n_2,m_2)$ represents
a $(n_1,m_1)$ inner tube concentrically aligned within an outer
$(n_2,m_2)$ tube, where $n_i$ and $m_i$ are the corresponding tube
indices. Monochiral DWNTs that have chiral walls with matching chiral
angles present axial facets like the achiral systems\cite{Leven2016}
and are therefore not considered herein. A summary of the relevant
geometric parameters of the unrelaxed DWNTs appears in Table~S2.

The structural and frictional properties of all DWNTs considered have
been described using dedicated intra- and inter-layer classical
force-fields as detailed below. For DWCNTs the intra-layer
interactions have been described using the Tersoff\cite{Tersoff}
potential adopting the parameterization of Lindsay and Broido.\cite{Lindsay}
The interlayer interactions of these systems have been described by the
registry-dependent Kolmogorov-Crespi potential in its {\small RDP1}
form.\cite{Kolmogorov2005} For the intra-layer interactions of DWBNNTs
we have used the Tersoff force-field as parameterized by Sevik et
al.\ for BN based systems,\cite{Cagin} along with our recently
developed {\it h}-BN interlayer potential with fixed partial
charges.\cite{Leven2014,Leven2017}
We note that suppressing the coulombic interactions between the
partially charged atomic centers in the DWBNNTs studied
($q_{\text{B}}$\,=\,+0.47\,$e$, $q_{\text{N}}$\,=\,-0.47\,$e$)
results in a reduction of merely $\sim$3.5\% in their calculated PES
corrugation (see Supporting Information Fig.~S3).
Corrugation and adhesion energy profiles for rigid planar bi-layer
of {\it h}-BN and graphene, as obtained by the above set of interlayer
potentials, are reported in Supporting Information Fig.~S4.

Periodic boundary conditions (PBC) along the tube axis have
been applied to all DWNTs considered, resulting in a very small
($<$0.1\%) stress in the case of the bi-chiral and mixed systems due to
the different lattice constants of the inner and outer tubes.
In all the cases, an initial step of relaxation of the cell vectors
has been performed in order to minimize any PBC-related stress effects.

In the pull-out/rotation potential energy surface calculations (Fig.~\ref{fig.PES})
each point has been obtained by placing the two unrelaxed cylindrical
nanotube walls at the corresponding relative axial and angular
position followed by geometry optimization using {\small FIRE}
quenched dynamics,\cite{FIRE} while nullifying the center of mass
(c.o.m.) axial and angular velocity of each nanotube wall.
All the constrained relaxations were stopped after 5000 FIRE iterations,
providing energy evaluations that are converged to within 0.1\% of the
highest energy obtained across the PES maps.
The reported energy per-atom has been obtained
by dividing the converged energy by the total number of atoms in the
DWNT. We note that this procedure corresponds to an adiabatic
relative motion of the tubes that can, in principle, be realized in
experiment by adhering the outer tube wall(s) to a fixed stage and
applying a slowly varying external force on the inner shells via the
manipulation of an external tip.\cite{Nigues2014,Ruoff2000a,Ruoff2000b,Akita2003}
Although in typical experimental setups the external force is applied
at one edge of the inner shells, their extreme stiffness permits the
instantaneous propagation of the stress along the entire tube length.
Hence, the calculated PESs should reliably describe the corresponding
inter-wall energy variations measured in the experiment.

Dynamic friction calculations have been performed by numerically
propagating the Langevin equation of motion using the standard
molecular dynamics velocity-Verlet algorithm.
The simulations have been performed in the underdamped regime by
applying viscous damping to all degrees of freedom apart from the
c.o.m.\ motion of both tubes. The dynamic friction force is evaluated
from the inter-wall shear force required to keep the two nanotube
walls at constant relative velocity motion $v_{ext}$. To this end, we
have fixed the c.o.m.\ of the internal tube and applied a uniform
force $F_{ext}$ to each of the $N$ atoms of the external tube so
that

\begin{align}
\begin{split}\label{eq.vi1}
  v_i^1 = v_i^0 + \left( \frac{ F_i+F_{ext} }{ m_i } - \gamma (v_i^0 - v_{cm}^0) \right) \Delta t
\end{split}\\
\begin{split}\label{eq.vi2}
  \sum_{i=1}^N v_i^1 = N v_{ext}
\end{split}
\end{align}

where $v_i^0$ and $v_i^1$ are the $i$-th atom velocities at times
$t_0$ and $t_1$, respectively, $\Delta t=t_1$\,-\,$t_0$ is the
numerical propagation timestep, $v_{cm}^0$ is the c.o.m.\ velocity of
the external tube at $t$\,=\,$t_0$, $m_i$ the atomic mass, $F_i$ is
the total force on atom $i$ due to the chosen set of interatomic
potentials, and $\gamma$\,=\,0.1\,ps$^{-1}$ is the viscous damping
coefficient used in the simulation to avoid system overheating. Since
the viscous damping is not applied to the c.o.m.\ motion of the tubes,
the computed friction results weakly dependent on the adopted $\gamma$
value, the latter mainly determining the steady-state temperature of
the sliding system. In our typical simulations, which were run in the
underdamped regime, we measured steady-state temperatures below 1\,K,
suggesting a negligible role of $\gamma$ on the measured friction.

From Eqs.~\ref{eq.vi1}-\ref{eq.vi2} we obtain
\begin{align}
\begin{split}\label{eq.Fext_i}
  F_{ext} = {}& \frac{\bar{m} v_{ext}}{\Delta t} + \bar{m}(\gamma - \frac{1}{\Delta t})\bar{v} \\
            {}& - \bar{m} \gamma v_{cm}^0 - \bar{m}\bar{a}~~,
\end{split}
\end{align}

where

\begin{align}
\begin{split}\label{eq.mbar}
  \bar{m} & = N\left(\sum_{i=1}^N \frac{1}{m_i} \right)^{-1}
\end{split}\\
\begin{split}\label{eq.abar}
  \bar{a} & = \frac{1}{N}\sum_{i=1}^N \frac{F_i}{m_i}
\end{split}\\
\begin{split}\label{eq.vbar}
  \bar{v} & = \frac{1}{N}\sum_{i=1}^N v_i^0
\end{split}
\end{align}

Since $F_{ext}$ is applied to all the atoms of the external tube, the
instantaneous friction force of the entire surface, $F_{fric}$, is simply
expressed by

\begin{align}
\begin{split}\label{eq.friction}
  F_{fric} = N F_{ext}
\end{split}
\end{align}

Finally, the obtained dynamic friction force
$F_{fric}$ is normalized to the inter-wall contact area evaluated from
the average diameter of the unrelaxed-cylindrical configuration
(see Table~S2), leading to the system-specific shear stress value.
This allows for a direct comparison among forces calculated for DWNTs
of different type and dimensions.
The resulting shear stress has been averaged over a time window of
at least 1\,ns during the steady-state motion, after the initial
transient dynamics has decayed, covering an integer number of oscillations
in the case of periodic force traces.

We note that using this procedure a direct quantitative comparison
with experimental data is hard to achieve, due to the large sliding
velocities, to which MD simulations are limited, compared to those
accessible in realistic experimental conditions. Despite this, our
dynamic simulations allow for a comparative study of the tribological
properties of faceted and unfaceted DWNTs of different chemical
composition.

\section*{Associated Content}

{\bf Supporting Information}

PES of armchair, zigzag, and bi-chiral DWCNTs (relaxed and cylindric); example of sinusoidal fitting of a PES to estimate static friction; comparison of the PES of armchair DWBNNT with and without partial charges contribution; corrugation and adhesion energy profiles for a bilayer of graphene and of {\it h}-BN; effect of sliding velocity on the average friction of armchair and bi-chiral DWCNTs and DWBNNTs; table reporting evaluated static friction and maximum corrugation energy for the reference DWNTs set; table reporting chiral angle difference, average diameter, and average interlayer spacing for the reference DWNTs set; movie of the cross sectional view of armchair and zigzag DWNTs relaxed at different relative axial positions z; movie showing a comparison of the cross sectional view of the armchair DWBNNT in telescopic motion at zero (adiabatic motion) and at large pulling velocity; movie of the bi-chiral DWBNNT at large relative pulling velocity; movie of the bi-chiral DWCNT at large relative pulling velocity.

\section*{Acknowledgments}
{\footnotesize Work in Trieste was carried out under ERC Grant 320796
  MODPHYSFRICT. EU COST Action MP1303 is also gratefully acknowledged.
  O.H. acknowledges the Lise-Meitner Minerva Center for Computational
  Quantum Chemistry and the Center for Nanoscience and Nanotechnology
  at Tel-Aviv University for their generous financial support.
}

\section*{Author contributions}
{\footnotesize R.G.\ and I.L.\ developed the simulation code and performed
  the calculations. R.G.\ conceived the method to evaluate dynamic friction,
  and produced the figures and the movies. All the authors actively
  participated in the data analysis and in the writing of the manuscript.
}

\end{document}